\documentclass[preprintnumbers,amsmath,amssymb,floatfix,11pt,prd,onecolumn,
superscriptaddress,nofootinbib]{revtex4-1}
\usepackage{latexsym}
\usepackage{epsfig}
\usepackage{epstopdf}
\usepackage{mathpazo}
\usepackage{amssymb}
\usepackage{mathtools}
\usepackage{amsmath,hyperref}
\usepackage{color}
\usepackage{graphicx}

\begin{document}

\title{Distinguishing rotating naked singularities from Kerr-like wormholes by their deflection angles of massive particles}

\author{Kimet Jusufi}
\email{kimet.jusufi@unite.edu.mk}
\affiliation{Physics Department, State University of Tetovo, Ilinden Street nn, 1200,
Tetovo, Macedonia.}
\affiliation{Institute of Physics, Faculty of Natural Sciences and Mathematics, Ss. Cyril and Methodius University, Arhimedova 3, 1000 Skopje, Macedonia.}
\author{ Ayan Banerjee}
\email{ayan\_7575@yahoo.co.in}
\affiliation{Astrophysics and Cosmology Research Unit, University of KwaZulu Natal, Private Bag X54001, Durban 4000,
South Africa.}

\author{Galin Gyulchev}

\email{gyulchev@phys.uni-sofia.bg}

\affiliation{Faculty of Physics, St. Kliment Ohridski University of Sofia, 5 James Bourchier Boulevard, Sofia 1164, Bulgaria}

\affiliation{Department of Physics, Biophysics and Roentgenology, Faculty of Medicine, St. Kliment Ohridski University of Sofia, 1, Kozyak Str., 1407 Sofia, Bulgaria.}

\author{Muhammed Amir}
\email{amirctp12@gmail.com}
\affiliation{Astrophysics and Cosmology Research Unit, University of KwaZulu Natal, Private Bag X54001, Durban 4000,
South Africa.}

\begin{abstract}
We study the gravitational deflection of relativistic massive particles by Janis-Newman-Winicour (JNW) spacetimes (also known as a rotating source with a surface-like naked singularity), and a rotating Kerr-like wormholes. Based on the recent article  [K. Jusufi, Phys. Rev. D 98, 064017 (2018)], we extend some of these results by exploring the effects of naked singularity and Kerr-like objects on the deflection of particles. We start by introducing coordinate transformation leading to an isotropic line element which gives the refraction index of light for the corresponding optical medias. On the other hand, the refraction index for massive particles is found by considering those particles as a de Broglie wave packets. To this end, we apply the Gauss-Bonnet theorem to the isotropic optical metrics to find the deflection angles. Our analysis shows that, in the case of  the JNW spacetime the deflection angle is affected by the parameter $0<\gamma<1$, similarly, we find that the deformation parameter $\lambda$ affects the deflection angle in the case of Kerr-like wormholes. In addition to that, we presented a detailed analysis of the deflection angle by means of the Hamilton-Jacobi equation that lead to the same results. As a special case of our results the deflection angle of light is recovered. Finally, we point out that the deflection of particles by Kerr-like wormholes is stronger compared to JNW spacetime, in particular this difference can be used to shed some light from observational point of view in order to distinguish the two spacetimes.

\end{abstract}


\date{\today}
\maketitle
\tableofcontents

\section{Introduction}

Deflection of light ray in weak gravitational field was the first experimental proof of the theory of general  
relativity for the Schwarzschild spacetime since 1919 \cite{Eddington}. Now it is serving as 
one of the important tool of probing a number of interesting phenomena in astronomy and cosmology. It is widely 
used for investigating the astrophysical objects, e.g., black holes, wormholes, super-dense neutron stars, 
naked singularity etc. This phenomenon has also been considered to study exoplanet detection from the dark
energy, dark matter, and Hubble parameter measurements. Moreover, the gravitational lensing 
(GL) provides an alternative approach towards mass measurement that leads to measure the mass 
of the white dwarf Stein 2051 B by using the Hubble space telescope \cite{Sahu}. Earlier works 
focused on the weak field theory of gravitational lensing which is based on the first order expansion of the 
small deflection angle \cite{Schneider}. Although the main disadvantage is when the lens is a 
very compact object (e.g. black hole), in this case the weak field approximation is no longer valid. 
The weak field approximation can not able to distinguish between various different solutions 
that are asymptotically flat. Therefore, 
after the pioneering work of Darwin \cite{Darwin}, the strong gravitational lensing has received much attentions. 
Note that in the case of strong lensing, the effect of lensing is strong, which means multiple imaging or 
strong deformation of the image. On the other hand, in case of weak lensing, the effect is weak, and the 
lensing images are characterized by weak distortions and small magnifications. In this case the fact of 
lensing cannot be identified by using an individual source, but can be revealed by statistical analysis of 
many objects. In addition, we can distinguish situations of lensing on the basis of the magnitude of the 
deflection angle: weak deflection of photons (with small deflection angles, sometimes referred as weak field 
situation) and strong deflection of photons (deflection angles are not small, sometimes referred as strong 
field situation), correspondingly. However, the crucial thing is that all observational areas of GL (strong, 
weak and micro) are based on using only weak deflection approximation meaning that, the deflection angle 
remains to be small. Strong deflection case on the other hand, is used for another problems: for example, for 
investigations of so-called relativistic images. To this end, we wish to point out the strong gravitational 
lensing for the Schwarzschild black hole \cite{Virbhadra:2007kw,Virbhadra:2008ws}, then extended to the 
Reissner-Nordstr{\"o}m black hole \cite{Eiroa:2002mk,Zhao:2015fya}, regular black holes \cite{Manna:2018jxb}, 
and wormholes \cite{Nandi:2006ds,Tsukamoto:2017hva,Shaikh:2017zfl,Tsukamoto:2017edq}, while the investigations 
of so-called relativistic images see also Refs. \cite{r1,r3,r4,r5,r6,r7}.

An interesting method has immersed for computing the angle of light deflection by Gibbons and Werner
\cite{GibbonsWerner1,GibbonsWerner2}. They considered the source and the receiver are located 
at an asymptotic region and then they applied the Gauss-Bonnet theorem (GBT) to the optical metric of a lens 
which allows us to describes a light ray as a spatial curve. Later on Werner \cite{Werner3} 
extended this approach to the stationary spacetime and computed the deflection angle for the Kerr black hole. 
This method has also been applied in the weak limit approximation to compute the deflection of 
light for global monopole and rotating cosmic string spacetime \cite{Jusufi:2017lsl,Jusufi:2017hed}. Note that the deflection angle of conical singularities has been pointed out in the old paper \cite{sokolov} . 
Nowadays the scientific community is discussing this phenomenon with great interest.
Along this line of thinking, Ishihara \emph{et al.} \cite{Ishihara:2016sfv} obtained the 
bending angle of light by using the GBT for finite distance from a lens object to a light source and a 
receiver. Further, this approach was extended by considering the stationary, axisymmetric and asymptotically 
flat spacetimes \cite{a2}. 
Jusufi and \"{O}vg\"{u}n \cite{kimet5} studied the effect of the cosmological constant on the 
deflection angle by a rotating cosmic string. For the influence of cosmological constant on the light deflection see the recent review  by Lebedev and Lake \cite{rcos}. On the other hand the GBT method has been considered for the several wormholes \cite{Jusufi:2017vta,Jusufi:2017mav,Rogatko:2018crz} and black holes 
\cite{Jusufi:2017drg,Jusufi:2018jof,Ovgun:2018prw} solutions. 

It is worth noticing that in the gravitational field the bending trajectories not only found 
for light or massless particles but also for massive particles. For a review on massive particles whose 
trajectories can go through a bending process, we refer the reader to \cite{Yu:2003hj,Brown:2005ta,Patla}, in particular the deflection angle for massive particles in weak
deflection case is calculated in Ref.   \cite{massweak}, and for strong deflection case see
\cite{massstrong1} and \cite{massstrong2}. 
Thus the important fact is trajectory bending due to the strong gravity near the black holes and the large 
velocity of the hypervelocity stars needs a full relativistic treatment, which will differ from the 
calculations made using only Newtonian gravity. In recent years, the existence of strong gravitationally lensed 
supernovae has long been predicted \cite{Holz,Goobar,Oguri,Oguri1}. Thus it is natural to expect that the 
neutrinos and cosmic rays emitted by these supernova and AGN are also lensed. Neutrino gravitational lensing 
by astrophysical objects has already been done in \cite{Escribano:2001ew}. Besides their positions, they also 
estimated the depletion of the neutrino flux after crossing a massive object \cite{Escribano:2001ew}. 
At the same time, Eiroa and Romero \cite{Eiroa:2008ks} have studied gravitational lensing of neutrinos by 
Schwarzschild black holes. 

Recently, Crisnejo and Gallo \cite{Crisnejo:2018uyn} showed that how one can successfully 
study the GBT theorem in a plasma medium for a static and spherically symmetric gravitational 
field. The gravitational deflection in plasma medium was considered in details 
before, see \cite{plasma1} and \cite{plasma2}. Even in the GBT method, they computed the deflection of massive particles with the same symmetries. 
On the other hand, Jusufi \cite{Jusufi:2018kry} introduced a new method based on GBT to 
compute the deflection angle for massive particles in a rotating spacetime geometry such as the Kerr black 
hole \cite{kerr} and Teo wormhole \cite{teo}. In doing so, he used an isotropic type metrics for a linearized 
rotating gravitational field where the refractive index of the corresponding optical media can be obtained. 
Motivated by the works we study the gravitational deflection of massive particles by Janis-Newman-Winicour 
(JNW) spacetimes \cite{Virbhadra:1997ie} (also known as a rotating source with a surface-like naked 
singularity), and a rotating Kerr-like wormholes \cite{Bueno:2017hyj}. Along this line of thought, recently 
different aspects of naked singularities and Kerr-like wormholes have been studied, including the iron line 
spectroscopy \cite{bambi}, accretion disc properties \cite{Joshi:2013dva}, spin precession of a test gyroscope 
\cite{Chakraborty:2016mhx,Rizwan:2018lht}, echoes of Kerr-like wormholes \cite{Bueno:2017hyj}, 
shadow images \cite{Amir:2018pcu} the strong and the weak deflection of light by Kerr-like wormholes 
\cite{Nandi:2018mzm,Ovgun:2018fnk}. In the present article, we study the deflection of massive particles 
by assuming the propagating particles as a de Broglie wave packets  introduced in \cite{nandi1}. 

The paper is organized as follows. In Sect.~\ref{defNak}, we use a coordinate transformation to recover 
an isotropic metric form for the rotating naked singularities, then derive the refractive 
index of the corresponding optical media. In Sect.~\ref{defkerr}, we consider the problem of the gravitational 
deflection of massive particles by Kerr-like wormholes. We present a detailed analyses of the 
deflection of massive particles by using the Hamilton-Jacobi approach in Sect.~\ref{geod}. We conclude our 
results in Sect.~\ref{concl}. We use the geometrized unit $G=c=\hbar= 1$ throughout this paper.

\section{Deflection of massive particles by Rotating Naked Singularities}
\label{defNak}
In this section, we calculate the deflection angle of rotating naked singularities thereby we consider
Janis-Newman-Winicour (JNW) spacetime. It is a solution of the Einstein equation described with the 
assumption of spherical symmetry, asymptotic flatness with the static massless scalar field $\Phi$. 
The action associated with this spacetime \cite{Virbhadra:1997ie,bambi} is
\begin{equation}
S=\int d^4x \sqrt{-g} \left[ R+g^{\mu \nu} (\partial_{\mu} \Phi) (\partial_{\nu} \Phi) \right],
\end{equation}
the variation of action turns out the following field equations
\begin{equation}
 R_{\mu \nu}= 8 \,(\partial_{\mu} \Phi) (\partial_{\nu} \Phi) , \quad \Box \Phi=0,
\end{equation}
where the scalar field is given by
\begin{equation}
\Phi=\frac{\sqrt{1-\gamma^2}}{4} \ln\left(1-\frac{2 mr }{\Sigma}\right).
\end{equation}
The corresponding rotating spacetime metric \cite{bambi} has the form
\begin{eqnarray}\label{3}\notag
ds^2 & = & \left(1-\frac{2 m r}{\Sigma}\right)^{\gamma}(dt-\omega d\phi)^2
+2 \omega (dt-\omega d\phi) d\phi \\
&-& \left(1 -\frac{2mr}{ \Sigma}\right)^{1-\gamma} 
\Sigma \left(\frac{dr^2}{\Delta} +d\theta^2 +\sin^2 \theta d\phi^2\right),
\end{eqnarray}
where $m=M/\gamma$, $M$ is the Arnowitt-Deser-Misner (ADM) mass, and
\begin{align}
\Sigma = r^2+a^2 \cos^2\theta, ~~\omega = a \sin^2 \theta, 
~~\Delta = r^2 +a^2-2mr, ~~\gamma = \frac{M}{\sqrt{M^2+q^2}}.
\end{align}
Note that $a =J/M$, where $J$ is the spin angular momentum of the source. For an instant 
if we set $\gamma=1$ and $q=0$, one can recover the Kerr black hole solution and the larger root of 
$\Delta = 0$ provides the radius of the event horizon, $r_{h}=M+\sqrt{M^2-a^2}$. It can be easily checked 
that for $0< \gamma <1$, and $q \neq 0$, the metric (\ref{3}) describes rotating naked singularities with 
a scalar curvature \cite{bambi}
\begin{equation}\label{sc}
 R=\frac{2(\gamma^2-1)m^2}{\Sigma^5}\left(1-\frac{2m r}{\Sigma}\right)^{\gamma-3}\left[\Delta (r^2- a^2 \cos^2 \theta)^2+(r a^2 \sin^2 (2 \theta))^2\right],
\end{equation}
and diverges on the surface
\begin{equation}
r_{\star}= m+\sqrt{m^2-a^2 cos^2\theta},
\end{equation}
$r_{\star}$ is the surface-like singularity, where curvature
invariants diverge and the spacetime is geodetically incomplete
as well.  In addition, the scalar curvature diverges where $g_{tt}=0$ and
represent the presence of a curvature singularity at
\begin{equation}
r_{\text{cs}}= \frac{1}{\gamma}\left[m+\sqrt{m^2-\gamma^2 a^2 cos^2\theta} \right],
\end{equation}
as the domain of variation of $\theta$ depends on the ratio $\gamma^2 a^2/m^2$. Also, it follows that 
in the transition from a black hole ($\gamma= 1$) to a naked singularity ($\gamma < 1$), the singularity 
is indeed naked, not the event horizon.

\subsection{Deflection angle}
Since, we are mainly interested in the weak limit, we shall simplify the problem by considering the 
deflection of particles in the equatorial plane ($\theta =\pi/2$) with a linearized metric in $a$, 
which results in
\begin{equation}
ds^2=\left[\left(1-\frac{2 m }{r}\right)^{\gamma}+2a \frac{d \phi}{dt} \left(1-\left(1-\frac{2m}{r}\right)^{\gamma}\right)\right]dt^2-\left(1-\frac{2m}{ r}\right)^{1-\gamma}\left(\frac{dr^2}{1-\frac{2m}{ r}}+r^2 d\phi^2\right).
\end{equation}
Now introducing a coordinate transformation via
\begin{equation}\label{eq10}
r=\rho \left(1+\frac{m}{2 \,\rho  }\right)^2,
\end{equation}
we deduce an isotropic metric with the line element given by
\begin{equation}
ds^2=\mathcal{F}^2(\rho) dt^2-\left(1+\frac{m}{2 \rho}\right)^4 \left(\frac{2 \rho-m}
{2 \rho+m}\right)^{2(1-\gamma)}\left(d\rho^2+\rho^2 d\varphi^2\right),
\end{equation}
where $\mathcal{F}^2(\rho)$ is
\begin{equation}
\mathcal{F}^2(\rho)=\left(\frac{2 \rho-m}{2 \rho+m}\right)^{2\gamma}+2a \frac{d \phi}{dt} \left(1-\left(\frac{2 \rho-m}{2 \rho+m}\right)^{2\gamma}\right).
\end{equation}
The velocity of light in this particular optical media can be found by setting $ds^2=0$, and using the 
definition $v=|d\vec{\rho}|/dt$, where $|d\vec{\rho}|^2=d\rho^2+\rho^2 d\varphi^2 $. With 
this information in hand, we can find the effective refractive index of the optical media by using 
$n(\rho)=c/v(\rho)$ and $c=1$, yielding
\begin{equation}
n(\rho)=\frac{\left(\frac{2 \rho+m}{2 \rho}\right)^2 \left(\frac{2 \rho-m}{2 \rho+m}\right)^{1-\gamma}}
{\sqrt{\left(\frac{2 \rho-m}{2 \rho+m}\right)^{2\gamma}-2 a \left(\frac{2 \rho-m}
{2 \rho+m}\right)^{2\gamma}\frac{d \varphi}{dt}+2 a \frac{d \varphi}{dt}}}.
\end{equation}
We can, therefore, write down the optical metric in terms of the refractive index as follows
\begin{equation}
dt^2=n(\rho)^2 d\rho^2+\rho^2 n(\rho)^2 d\varphi^2.
\end{equation}
By inverting the coordinate transformation \eqref{eq10}, we obtain
\begin{equation}\label{eq15}
\rho = \frac{r-m+\sqrt{r^2-2m r}}{2},
\end{equation} 
which shows leading terms in $\rho \simeq r - m$. The above relations holds for the deflection of light. 
However, we are interested in studying the deflection of massive particles. One way to incorporate this 
is to consider the propagating massive particles as the de Broglie wave packets. In particular, following 
\cite{Jusufi:2018kry} and \cite{nandi1}, we can argue that by using $p=\hbar k$ and $H=E=\hbar \omega $, 
(note that we have temporarily introduced $\hbar$) results
\begin{equation}
\lambda=\frac{h}{n H \sqrt{1-\frac{\mu^2 \mathcal{F}^2 }{H^2}}},
\end{equation}
where $\mu$ is the rest mass of the particle. We rearrange this equation to obtain a constant quantity 
of the given optical media
\begin{equation}
\lambda n \sqrt{1-\frac{\mu^2 \mathcal{F}^2 }{H^2}}=\frac{h}{ H }=const.
\end{equation}
In other words, this equation is a generalization of well-known result in wave-optics. For massive 
particles, in a given isotropic metric, the expression should be constant everywhere in the optical 
medium, i.e., $\lambda N=const$. Thus the refractive index relation for massive particles is given by
\begin{equation}
n(\rho)_{mass. \,particles} \to N(\rho)=n(\rho)\sqrt{1-\frac{\mu^2}{E^2}\mathcal{F}^2(\rho)},
\end{equation}
modifying the optical metric
\begin{equation}\label{eq19}
\mathrm{d}\sigma^2_{mass. particles}=N(\rho)^2 \mathrm{d}\rho^2+\rho^2 N(\rho)^2 \mathrm{d}\varphi^2.
\end{equation}
Furthermore, the energy of the particles measured at infinity is given by
\begin{equation}\label{Energy}
E =\frac{\mu }{\left(1-w^2\right)^{1/2}},
\end{equation}
where $w$ represents the relativistic velocity. The angular momentum can be written as
\begin{equation}\label{AngMomentum}
J=\frac{\mu w b}{\left(1-w^2\right)^{1/2}},
\end{equation}
where $b$ is the impact parameter. Following the definition of the impact parameter, we can write
\begin{equation}
\frac{J}{E }=w\,b,
\end{equation}
which reduces to $b$ in the case of light $w=c=1$. Therefore, the quantity  $\mathrm{d} \varphi /\mathrm{d}t$, in case of massive particles is modified as follow
\begin{equation}
\frac{\mathrm{d} \varphi}{\mathrm{d}t}=\frac{a+\left(1-\frac{2m}{ r}\right)^{\gamma} b w-(1-\frac{2m}{ r})^{\gamma}a}{\left(1-\frac{2m}{ r}\right)^{1-\gamma}r^2+\left(1-\frac{2m}{ r}\right)^{\gamma} a b w},
\end{equation}
yielding
\begin{equation}
N(r)=w+\frac{\gamma m (r^2w^2-2 ab w+r^2)}{w r^3}+\mathcal{O}(m^2,a^2).\label{eq24}
\end{equation}
This result clearly indicates that the refractive index is modified due to the angular momentum 
parameter $a$. 

Let us continue to calculate the Gaussian optical curvature using metric \eqref{eq19}. Rewriting the optical metric \eqref{eq19}, in terms of a new coordinates,
\begin{eqnarray}
\mathrm{d}r^{\star}=N(\rho)\, \mathrm{d}\rho,\,\,\,f(r^{\star})=N(\rho)\, \rho,
\end{eqnarray}
thus the optical metric reads
\begin{eqnarray}
\mathrm{d}\sigma^{2}=\tilde{g}_{ab}\,\mathrm{d}x^{a}\mathrm{d}x^{b}=\mathrm{d}{%
r^{\star }}^{2}+f^{2}(r^{\star })\mathrm{d}\varphi ^{2}. 
\end{eqnarray}
In terms of these coordinates, the Gaussian optical curvature $\mathcal{K}$ can be expressed as: 
\begin{eqnarray}
  \mathcal{K} & = & - \frac{1}{f (r^{\star})}  \frac{\mathrm{d}^2 f
  (r^{\star})}{\mathrm{d} r^{\star 2}} \\\notag
  & = & - \frac{1}{f (r^{\star})}  \left[ \frac{\mathrm{d} \rho}{\mathrm{d}
  r^{\star}}  \frac{\mathrm{d}}{\mathrm{d} \rho} \left( \frac{\mathrm{d}
  \rho}{\mathrm{d} r^{\star}} \right) \frac{\mathrm{d} f}{\mathrm{d} \rho} + \left(
  \frac{\mathrm{d} \rho}{\mathrm{d} r^{\star}} \right)^2 \frac{\mathrm{d}^2
  f}{\mathrm{d} \rho^2} \right]. 
\end{eqnarray}
We can, therefore, express the Gaussian optical curvature for massive particles in terms of the modified refractive index as
\begin{align}
\begin{split}
\mathcal{K} = & -\frac{N(\rho) N''(\rho) \rho -(N'(\rho))^2\rho+N(\rho) N'(\rho)}{N^4(\rho) \rho}.
\end{split}
\end{align} 
In particular, by using \eqref{eq15} we can rewrite the last equation, in terms of the old coordinate 
$r$, which in leading order terms results with
\begin{equation}
\mathcal{K} \simeq - \frac{m \gamma (1+w^2)}{r^3 w^4}+\frac{18 m \gamma b a}{r^5 w^3}+\mathcal{O}(m^2,a^2).
\end{equation}
With these results in hand, we can chose a non-singular region $\mathcal{D}_{R}$ with boundary $\partial
\mathcal{D}_{R}=\gamma _{\tilde{g}}\cup C_{R}$. The GBT provides a connection between geometry (in a sense of optical curvature) and topology (in a sense of Euler characteristic number) stated as follow
\begin{equation}
\iint\limits_{\mathcal{D}_{R}}\mathcal{K}\,\mathrm{d}S+\oint\limits_{\partial \mathcal{%
D}_{R}}\kappa \,\mathrm{d}\sigma+\sum_{i}\theta _{i}=2\pi \chi (\mathcal{D}_{R}),
\end{equation}
with $\kappa$ is the geodesic curvature, and $\mathcal{K}$ is the Gaussian optical curvature. 
Now we can choose a non-singular domain with Euler characteristic number 
$\chi (\mathcal{D}_{R})=1$. The geodesic curvature $\kappa$, in the case of massive particles is 
defined by the relation
\begin{equation}
\kappa =\tilde{g}\,\left(\nabla _{\dot{\gamma}}\dot{\gamma},\ddot{\gamma}\right),
\end{equation}
but also keeping in mind an additional unit speed condition $\tilde{g}(\dot{\gamma},\dot{%
\gamma})=1$, with $\ddot{\gamma}$ being the unit acceleration vector. The two corresponding jump angles in the limit $R\rightarrow \infty $, reads  $\theta _{\mathit{O}%
}+\theta _{\mathit{S}}\rightarrow \pi $. Therefore the GBT now is simplified as follow
\begin{equation}
\lim_{R\rightarrow \infty }\int_{0}^{\pi+\hat{\alpha}}\left[\kappa \frac{\mathrm{d} \sigma}{\mathrm{d} \varphi }\right]_{C_R} \mathrm{d} \varphi =\pi-\lim_{R\rightarrow \infty }\iint\limits_{\mathcal{D}_{R}}\mathcal{K}\,\mathrm{d}S.
\end{equation}
By construction, there is a zero contribution from the geodesics i.e. $\kappa (\gamma_{\tilde{g}})=0$, therefore, we end up with a contribution only for the the curve $C_{R}$, thus we can write 
\begin{equation}
\kappa (C_{R})=|\nabla _{\dot{C}_{R}}\dot{C}_{R}|.
\end{equation}
To evaluate the above expression one can express the curve in terms of the radial coordinate 
$C_{R}:=r(\varphi)=R=\text{const}$, resulting with a nonzero contribution for the radial component
\begin{equation}
\left( \nabla _{\dot{C}_{R}}\dot{C}_{R}\right) ^{r}=\dot{C}_{R}^{\varphi
}\,\left( \partial _{\varphi }\dot{C}_{R}^{r}\right) +\tilde{\Gamma} _{\varphi
\varphi }^{r}\left( \dot{C}_{R}^{\varphi }\right) ^{2}. 
\end{equation}
The first term vanishes, however, using the unit speed condition and the optical metric, we find the 
final result 
\begin{eqnarray}\notag
\lim_{R\rightarrow \infty }\kappa (C_{R}) &=&\lim_{R\rightarrow \infty
}\left\vert \nabla _{\dot{C}_{R}}\dot{C}_{R}\right\vert , \notag \\
&\rightarrow &\frac{1}{w R}. 
\end{eqnarray}
We see that the final result of the geodesic curvature is slightly modified, letting $w=1$, we recover $\kappa (C_{R}) \to R^{-1}$, as expected.  For an observer located at the coordinate $R$, we find 
\begin{eqnarray}
\lim_{R\rightarrow \infty } d \sigma &\to & w R \, d\varphi.  
\end{eqnarray}
Putting together these results, from the GBT, we find the following expression for the deflection angle  
\begin{equation}
\hat{\alpha}=-\int\limits_{0}^{\pi}\int\limits_{\frac{b}{\sin \varphi}}^{\infty}\left(- \frac{m \gamma (1+w^2)}{r^3 w^4}+\frac{18 m \gamma b a}{r^5 w^3}\right)dS,
\end{equation}
where the surface element is given by $dS=\sqrt{\tilde{g}} \,dr^{\star} d\varphi \simeq N^2(r) r dr d\varphi $. In addition to that, from Eq. \eqref{eq24} the following approximation holds $N(r)^2 \simeq w^2$. This integral can easily be evaluated,
\begin{equation}\label{DARNSGBT}
\hat{\alpha}\simeq \frac{2 m \gamma}{b w^2}(1+w^2) \pm \frac{4 m a \gamma}{b^2 w},
\end{equation}
where the plus and minus sign refers to the retrograde and prograde light rays, respectively. 
That being said, it is clear that the deflection angle is bigger in the case of retrograde orbit, and smaller in 
the case of prograde orbit. Now we can consider three special cases in the last equation: 
(i) If $\gamma =1$, we find the deflection angle of massive particles in Kerr black hole geometry. (ii) If 
$w=1$, then the deflection angle of light by a JNW spacetime is recovered.  (iii) Finally, 
letting $\gamma=w=1$, we recover the deflection of light in a Kerr geometry. Our result for the deflection 
angle is consistent with previous studies 
\cite{weaklens1,weaklens2,weaklens3,weaklens4,weaklens5,weaklens6,weaklens7,weaklens8}.  It is important to 
realize that our method applies only for relativistic particles, i.e. $0<w \leq 1$, in other words, there is an 
apparent singularity in the deflection angle in the limit $w \to 0$. Of course, this apparent singularity can 
be removed but a more general setup is needed. 
\begin{figure}[h!]
\includegraphics[width=0.45\textwidth]{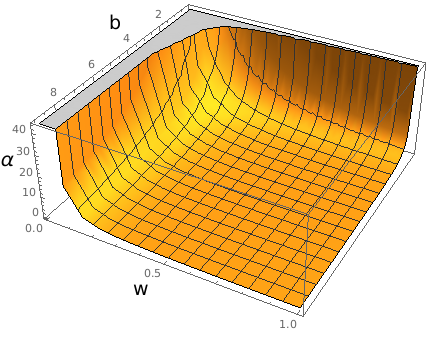} \includegraphics[width=0.45\textwidth]{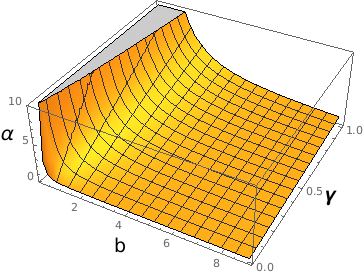}
\caption{In the left panel we have plotted the deflection angle as a function of $b$ and $\gamma$ when $w=0.5$, $m=a=1$, and in the right panel same thing plotted as a function of $b$ and $w$ for $\gamma=0.5$, $m=a=1$.}
\label{w1}
\end{figure}

\section{Deflection of massive particles by Kerr-like wormholes}
\label{defkerr}
We consider a Kerr-like wormhole spacetime, recently proposed in \cite{Bueno:2017hyj}. The spacetime 
metric can be written, in Boyer-Lindquist coordinates ($t, r, \theta, \phi$) as 
\begin{eqnarray}\label{metric}
ds^2 & = & \left(1-\frac{2Mr}{\Sigma}\right)dt^2 +\frac{4Mar\sin^2 \theta}{\Sigma} dt d\phi 
-\frac{\Sigma}{\hat{\Delta}}dr^2   \nonumber \\ &-& \Sigma d \theta^2 
- \left(r^2+ a^2 +\frac{2Ma^2r \sin^2 \theta}{\Sigma} \right) \sin^2 \theta d\phi^2,
\end{eqnarray}
where $\Sigma$ and $\hat{\Delta}$ are expressed by
\begin{eqnarray}
\Sigma = r^2 + a^2 \cos^2\theta, \quad \hat{\Delta}=r^2 + a^2 - 2 M(1+\lambda^2)r.
\end{eqnarray}
It contains a family of parameters where $a$ and $M$ corresponds to the spin and the mass of wormhole, 
and $\lambda^2$ is the deformation parameter. A non vanishing $\lambda^2$ differs this metric from Kerr 
metric, we can recover the Kerr metric when $\lambda^2 =0$. Although the throat of the wormhole can be 
easily obtained by equating the $\hat{\Delta}$ to zero,
\begin{equation}
r_{+} = (1+\lambda^2) M + \sqrt{M^2(1+\lambda^2)^2 -a^2},
\end{equation}
which represents a special region that connects two different asymptotically flat regions. To simplify 
the problem, we can consider the deflection in the equatorial plane in a linearized Kerr-like wormhole 
metric in spin $a$. Such a metric can also describe the gravitational field around the rotating 
star or planet given by 
\begin{equation}
ds^2 \simeq \left(1-\frac{2M}{r}+\frac{4Ma  }{r} \frac{d\varphi}{dt}\right)dt^2
- \left(1-\frac{2M(1+\lambda^2)}{r}\right)^{-1} dr^2 -r^2d\varphi^2.
\end{equation}

\subsection{Deflection angle}
We follow the same GBT approach to calculate the deflection angle of Kerr-like wormhole as discussed 
in previous section. Here we use the following coordinate transformation 
\begin{equation}\label{eq43}
r=\rho \left(1+\frac{M(1+\lambda^2)}{2 \rho}\right)^2,
\end{equation}
after substitution the metric takes the following form 
\begin{equation}
ds^2=\frac{\left(1+\frac{M(1+\lambda^2)}{2\rho}\right)^2-\frac{2M}{\rho}+\frac{4Ma}{\rho}\frac{d\varphi}
{dt}}{\left(1+\frac{M (1+\lambda^2)}{2\rho}\right)^2}dt^2-\left(1+\frac{M (1+\lambda^2)}
{2\rho}\right)^4\left(d\rho^2+\rho^2 d\varphi^2\right).
\end{equation}
The last equation represents an isotropic metric, so that the line element can be written in the form
\begin{equation}
ds^2=\mathcal{F}^2(\rho) dt^2-\mathcal{G}^2(\rho) |d\vec{\rho}|^2.
\end{equation}
The isotropic coordinate speed of light $v(\rho)$ can be found from the relation
\begin{equation}
v(\rho)=|\frac{d\vec{\rho}}{dt}|=\frac{\left[\left(1+\frac{M(1+\lambda^2)}{2\rho}\right)^2-\frac{2M}{\rho}+\frac{4Ma}{\rho}\frac{d\varphi}{dt}\right]^{1/2}}{\left(1+\frac{M (1+\lambda^2)}{2\rho}\right)^3}.
\end{equation}
Using $n(\rho)=c/v(\rho)$ and $c=1$, the last equation yields the effective refractive index for light 
in the Kerr gravitational field 
\begin{equation}
n(\rho)=\frac{\left(1+\frac{M (1+\lambda^2)}{2\rho}\right)^3}{\left[\left(1+\frac{M(1+\lambda^2)}
{2\rho}\right)^2-\frac{2M}{\rho}+\frac{4Ma}{\rho}\frac{d\varphi}{dt}\right]^{1/2}},
\end{equation}
In this way the optical metric reads
\begin{equation}
dt^2=n(\rho)^2 d\rho^2+\rho^2 n(\rho)^2 d\varphi^2.
\end{equation}
By inverting the coordinate relation \eqref{eq43}, we find
\begin{equation}
\rho = \frac{1}{2}\left[r-M(1+\lambda^2)\pm \sqrt{r^2-2Mr(1+\lambda^2)}\right],
\end{equation} 
which suggests that in leading order terms $\rho \simeq r -M(1+\lambda^2)$. The refractive index 
relation for massive particles in that case reads
\begin{equation}
N(r)=n(r)\sqrt{1-\frac{m^2}{E^2}\mathcal{F}^2(r)}.
\end{equation}
We can apply this expression to our Kerr optical media which results with
\begin{equation}
\mathcal{F}^2(r)=\frac{\left(1+\frac{M(1+\lambda^2)}{2 r}\right)^2-\frac{2M}{r}
+\frac{4Ma}{r}\frac{d\varphi}{dt}}{\left(1+\frac{M (1+\lambda^2)}{2r}\right)^2},
\end{equation}
in which Eq. \eqref{eq43} has been used. 
Therefore, the quantity  $d \varphi /dt$, in the case of massive particles is modified as follow
\begin{equation}
\frac{d \varphi}{dt}=\frac{2M a+(r-2M) w b}{r^3-2Ma w b},
\end{equation}
yielding
\begin{equation}
N(r)=w+\frac{w \left(r^2 \lambda^2 w^2+r^2 w^2-2 a b w+r^2 \right)M}{r^3 w^2}+\mathcal{O}(M^2,a^2).
\end{equation}
This result clearly indicates that the refractive index is modified due to the angular momentum parameter $a$. In particular, in leading order terms, we find
\begin{equation}
\mathcal{K} \simeq - \frac{\left(w^2 r^2-18 ab w+r^2\right)M}{w^4 r^5}
-\frac{\lambda^2 M}{r^3 w^3 }+\mathcal{O}(M^2,a^2).
\end{equation} 
%
%
%
Going through the same procedure we find
\begin{eqnarray}\notag
\lim_{R\rightarrow \infty }\kappa (C_{R}) &=&\lim_{R\rightarrow \infty
}\left\vert \nabla _{\dot{C}_{R}}\dot{C}_{R}\right\vert , \notag \\
&\rightarrow &\frac{1}{w R}. 
\end{eqnarray}
From the optical metric is quite easy to see that at a very large distance 
\begin{eqnarray}
\lim_{R\rightarrow \infty } d\sigma &\to & w R \, d\varphi.  
\end{eqnarray}
Now from the GBT we find the following expression for the deflection angle  
\begin{equation}
\hat{\alpha}=-\int\limits_{0}^{\pi}\int\limits_{\frac{b}{\sin \varphi}}^{\infty}\left(- \frac{\left(w^2 r^2-18 ab w+r^2\right)M}{w^4 r^5}-\frac{\lambda^2 M}{r^3 w^3 }\right) w^2 r dr d\varphi.
\end{equation}
This integral can easily be evaluated, yielding
\begin{equation}\label{DAKLWGBT}
\hat{\alpha}\simeq \frac{2M}{b}\left(1+\lambda^2+\frac{1}{w^2}\right)\pm \frac{4Ma}{b^2}\frac{1}{w}.
\end{equation}
Note that the plus and minus signs refers to the retrograde and prograde light ray, 
respectively. Setting $w=1$, in the last equation we recover the deflection angle of light.
\begin{figure}[h!]
\includegraphics[width=0.45\textwidth]{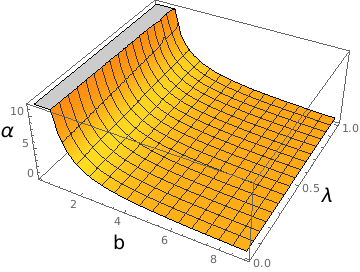}\includegraphics[width=0.45\textwidth]{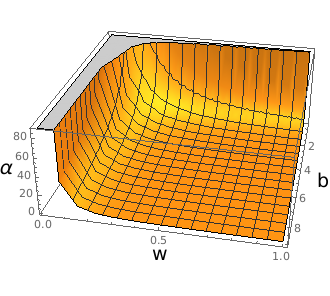}
\caption{The deflection angle has been plotted as a function of $b$ and $\lambda$ when $w=0.8$, $M=a=1$ in the left panel, whereas for a function of $b$ and $w$ when $\lambda=0.1$, $M=a=1$ in the right panel.}
\label{w2}
\end{figure}

\section{Geodesics Approach}
\label{geod}
In this section, we will derive the equations of motion of an uncharged test particle of mass $\mu$ 
and then we will calculate the deflection angle of the particle moving in the rotating naked 
singularity or the Kerr-like wormhole spacetime, given by (\ref{3}) and (\ref{metric}), respectively. 
The geodesic equations may be derived using Hamilton-Jacobi approach, as we look for a function of 
the coordinates and the geodesic affine parameter $\sigma$,
\begin{equation}\label{WaveFrontFunction}
    S=S(x^{\mu},\sigma),
\end{equation}
which is a solution of the Hamilton-Jacobi equation
\begin{equation}\label{HJEq}
    \mathcal{H}\left(x^{\mu},\frac{\partial S}{\partial x^{\mu}}\right)
    +\frac{\partial S}{\partial \sigma}=0.
\end{equation}
In general, the solution $S$ of the Hamiltonian-Jacobi equation depends on four integration constants and satisfy the relations for the conjugate momenta
\begin{equation}\label{Impulses}
    \frac{\partial S}{\partial x^{\mu}}=p_{\mu}(\sigma),
\end{equation}
which can be defined by the Lagrangian of the system 
$\mathcal{L}=\frac{1}{2}g_{\mu\nu}\dot{x}^{\mu}\dot{x}^{\nu}$ through the equations 
$p_{\mu}-\partial \mathcal{L}/\partial \dot{x}^{\mu}=0$. Here, the dot represents differentiation with 
respect to the geodesic affine parameter $\sigma$. In order to involve the mass of the test particle, 
$\sigma$ must be related to the proper time by $\tau=\mu \sigma$, which turns the conjugate momenta 
into
\begin{equation}\label{Impulses}
    p_{\mu}(\tau)=\mu g_{\mu\nu}u^{\nu},
\end{equation}
where $u^{\mu}\equiv d x^{\mu}/d\sigma$ stands for the four-velocity of the particle with normalization 
condition $g_{\sigma\rho}u^{\sigma}u^{\rho}=1$. Therefore, one can obtain the Hamiltonian
\begin{equation}\label{Hamiltonian}
    \mathcal{H}=\frac{1}{2}g^{\mu\nu}p_{\mu}p_{\nu}=\frac{1}{2}\mu^2,
\end{equation}
which does not depend explicitly on geodesic affine parameter $\sigma$ and is a constant of motion. By 
taking zero or positive values of $\mu^2$, the Hamiltonian (\ref{Hamiltonian}) can be used to give null 
or timelike geodesics, respectively.
The corresponding Hamilton-Jacobi equation is
\begin{equation}\label{HJE}
    -\frac{\partial S}{\partial \sigma}=\frac{1}{2}g^{\mu\nu}\frac{\partial S}{\partial x^{\mu}}
    \frac{\partial S}{\partial x^{\nu}},
\end{equation}
where $S$ is the Jacobi function (\ref{WaveFrontFunction}). From the symmetries one can obtain two constants of motion corresponding to conservation of energy, $E$, and angular momentum with respect to the rotational symmetry axis of the naked singularity or Kerr-like wormhole, $J$. Therefore, we have
\begin{equation}\label{ConservationEL}
\begin{split}
    p_{t}&=g_{tt}\dot{t}+g_{t\phi}\dot{\phi}=E, \\
    p_{\phi}&=g_{\phi\phi}\dot{\phi}+g_{\phi t}\dot{t}=-J.
\end{split}
\end{equation}
It follows that for the time and azimuthal coordinates we obtain the equations
\begin{equation}\label{TimeAzimuthalEq}
\begin{split}
    \dot{t}&=-\frac{g_{\phi\phi}\,E+g_{t\phi}\,J}{g_{t\phi}^2-g_{tt}g_{\phi\phi}}, \\
    \dot{\phi}&=\frac{g_{t\phi}\,E+g_{tt}\,J}{g_{t\phi}^2-g_{tt}g_{\phi\phi}}.
\end{split}
\end{equation}
In addition, we have the constant of motion given by Eq. (\ref{Hamiltonian}), corresponding to the 
conservation of the rest mass of the particle. The energy $E$ and the angular momentum of the particle 
$J$ can be expressed by relativistic velocity $w$ and impact parameter $b$ at infinity, where the 
spacetime is asymptotically flat, as it is shown by Eqs. (\ref{Energy}) and (\ref{AngMomentum}).

Hereafter, we will consider only a light ray laying on $\theta=\pi/2$ hypersurface of the considered 
rotating naked singularity or wormhole geometry expecting the largest ray deviation. Therefore, we 
shall seek a solution of the Hamilton-Jacobi equation of the form
\begin{equation}\label{SolutionForm}
    S=-\frac{1}{2}\mu^2 \sigma +E\,t -J\,\phi +S_{r}(r,E,J).
\end{equation}
Taking into account the relationship between a metric tensor and its inverse, thus the nonzero 
contravariant metric coefficients are
\begin{equation}\label{InvMCoeff}
\begin{split}
    g^{tt} & =\frac{g_{\phi\phi}}{g_{tt}g_{\phi\phi}-g_{t\phi}^2}, \hspace{0.5cm} g^{rr}=\frac{1}{g_{rr}}, \hspace{0.5cm} g^{\theta\theta}=\frac{1}{g_{\theta\theta}}, \\
    g^{\phi\phi} & =\frac{g_{tt}}{g_{tt}g_{\phi\phi}-g_{t\phi}^2}, \hspace{0.5cm} g^{t\phi}=g^{\phi t}=-\frac{g_{t\phi}}{g_{tt}g_{\phi\phi}-g_{t\phi}^2}.
\end{split}
\end{equation}
The Hamilton-Jacobi equation than takes the form
\begin{equation}\label{HJE2}
    \frac{1}{g_{rr}}\left(\frac{d S_{r}(r)}{d r}\right)^2+\frac{g_{\phi\phi}E^2+2g_{t\phi}EJ+g_{tt}J^2}{g_{tt}g_{\phi\phi}-g_{t\phi}^2}=\mu^2.
\end{equation}
Using the relation $p_{r}=g_{rr}\dot{r}=\partial S/\partial r=dS_{r}/dr$, it may be related directly to the radial equation of motion of a particle with rest mass $\mu$
\begin{equation}\label{RadialEqMotion}
\begin{split}
    \left(\frac{dr}{d\sigma}\right)^2 & =\frac{g_{\phi\phi}E^2+2g_{t\phi}EJ+g_{tt}J^2}{g_{rr}(g_{t\phi}^2-g_{tt}g_{\phi\phi})}+\frac{\mu^2}{g_{rr}} \\
              & =\frac{g_{\phi\phi}(E-V_{+})(E-V_{-})}{g_{rr}(g_{t\phi}^2-g_{tt}g_{\phi\phi})}+\frac{\mu^2}{g_{rr}},
\end{split}
\end{equation}
where the solutions of the equation in $E$
\begin{equation}\label{EffPotentialEq}
    g_{\phi\phi}E^2+2g_{t\phi}EJ+g_{tt}J^2=0
\end{equation}
represent the effective potentials
\begin{equation}\label{EffPotensials}
    V_{\pm}(r)=\frac{J}{g_{\phi\phi}}\bigg(-g_{t\phi}\pm\sqrt{g_{t\phi}^2-g_{tt}g_{\phi\phi}}\bigg).
\end{equation}
Evaluating the Hamiltonian at the minimum distance of approach of the particle, where $\dot{r}$ = 0, we find the precise relation between the impact parameter $\xi=J/E=wb$ and the closest approach distance $r_{0}$
\begin{equation}\label{ImpactParameterEq}
    \xi(r_0)=\frac{-g_{t\phi}+\sqrt{(g_{t\phi}^2-g_{tt}g_{\phi\phi})[1-g_{tt}(1-w^2)]}}{g_{tt}}\Big{|}_{r_{0}},
\end{equation}
which reduces to expected photon impact parameter $b$ for stationary axially symmetric metrics, when $w=c=1$ \cite{Bozza}. The sign in front of the square root is chosen to be positive when the orbital motion of the particle and rotation of the compact object are in the same direction.

The orbital equation of the particle in terms of azimuthal $\phi$ and radial $r$ coordinates can be obtained by using the fact that the partial derivative of the Jacobi function with respect to the constant of motion $J$ is a constant
\begin{equation}\label{EqMotionPhir}
    \frac{\partial S}{\partial J}=\phi-\frac{\partial S_{r}(r,E,J)}{\partial J}=const.
\end{equation}
Solving Eq. (\ref{HJE2}) or taking into account Eq. (\ref{RadialEqMotion}), we obtain the azimuthal shift function of the particle by Eq. (\ref{EqMotionPhir})
\begin{equation}\label{OrbitalEq}
\begin{split}
    \phi(r)-\phi(r_{0}) & =\frac{\partial S_{r}(r,E,J)}{\partial J} \\
        & =\quad \int_{r_{0}}^{r}g_{rr}\frac{\partial}{\partial J}\left(\frac{dr}{d\sigma}\right)dr \\
        & =\pm\int_{r_{0}}^{r}\frac{(g_{t\phi}E+g_{tt}J)}{(g_{t\phi}^2-g_{tt}g_{\phi\phi})}  \left[\frac{g_{\phi\phi}(E-V_{+})(E-V_{-})}{g_{rr}(g_{t\phi}^2-g_{tt}g_{\phi\phi})}+\frac{\mu^2}{g_{rr}}\right]^{-1/2}dr \\
            & =\pm\int_{r_{0}}^{r}\frac{(g_{t\phi}+\xi g_{tt})}{(g_{t\phi}^2-g_{tt}g_{\phi\phi})}  \left[\frac{g_{\phi\phi}(1-\mathcal{V}_{+})(1-\mathcal{V}_{-})}{g_{rr}(g_{t\phi}^2-g_{tt}g_{\phi\phi})}+\frac{1-w^2}{g_{rr}}\right]^{-1/2}dr,
\end{split}
\end{equation}
where the sign in front of the integral is changing when the particle passes thought a turning point $r_{0}$ of the orbital motion. $\mathcal{V}_{\pm}(r,\xi)\equiv V_{\pm}(r,J)/E$ stand for the effective potentials per unit of the energy and depend directly on the impact parameter $\xi$. It can be verified that an Eq. (\ref{OrbitalEq}) is consistent with Eqs. (\ref{TimeAzimuthalEq}) and (\ref{RadialEqMotion}).

Having the azimuthal shift of the photon as a function of the coordinate distance $r$, we can calculate the deflection angle of the test-particle as we assume that the source $r_{S}$ and observer $r_{O}$ are placed in the asymptotically flat region of the spacetime, so that $r_{S}\rightarrow\infty$ and $r_{O}\rightarrow\infty$. Then, after calculating the total azimuthal angle according to the orbital equation (\ref{OrbitalEq}), the deflection angle is
\begin{eqnarray}\label{DA}
    \hat{\alpha}&=&\phi(r_{S})-\phi(r_{O})-\pi \\
                &=&2|\phi(r_{\infty})-\phi(r_{0})|-\pi.
\end{eqnarray}

\subsection{Deflection angle by Rotating Naked Singularities}
In order to calculate the deflection angle of a test-particle in the spacetime of rotating naked singularities, firstly we will calculate the impact parameter of the test-particle $\xi=wb$ as a function of the distance of closest approach $r_{0}$. According to Eq. (\ref{ImpactParameterEq}) the equatorial impact parameter of a particle with relativistic velocity $w$ is given by
\begin{equation}\label{IParameterRNS}
    \xi(r_{0})=\frac{  a\left(1-\frac{2m}{r_{0}}\right)^{\gamma}+a+\sqrt{(r_{0}^2+a^2-2mr_{0})\left[\left(1-\frac{2m}{r_{0}}\right)^{\gamma}(w^2-1)+1\right]} }{\left(1-\frac{2m}{r_{0}}\right)^{\gamma}}
\end{equation}
The impact parameter (\ref{IParameterRNS}) reduces to that for photons, obtained in \cite{Gyulchev1}, letting $w = 1$. For $a > 0$ the naked singularity rotates counterclockwise, while for $a < 0$ the naked singularity and the particles rotate in converse direction.

Expecting small deviations in the orbital motion of the test-particles, we will assume that the distance of closest approach $r_{0}$ is of the same order as the impact parameter $b$ due to the validity of inequalities, $b\gg m$ and $b\gg a$. Therefore, considering Eq. (\ref{RadialEqMotion}) in the framework of the assumptions adopted, we suggest the following solution
\begin{equation}\label{r0approximation1}
    r_{0}\simeq b \left\{1+\sum\limits_{i,j=1}^{2} c_{m,a}\,\epsilon_{m}^{i}\epsilon_{a}^{j}+\mathcal{O}(\epsilon^3)\right\},
\end{equation}
where we are introducing two independent expansion parameters, as the former is in term of metric parameter $m$ related to the ADM mass $M$ and scalar charge $q$, while the latter is in term of specific angular momentum $a$ of the rotating naked singularity
\begin{equation}\label{epsma}
    \epsilon_{m}=\frac{m}{b}, \quad\quad \epsilon_{a}=\frac{a}{b}.
\end{equation}
Here the coefficients $c_{m,a}$ are real numbers, and the summation is over all possible combinations of epsilon powers $i, j$ up to and including second order terms.

Applying the procedure to the equation $\dot{r}(r_{0})=0$ as discussed above, we acquire the following result
\begin{equation}\label{ClosestDistanceRNS}
\begin{split}
    r_{0}\simeq b & \left\{1+\frac{[w^2-\gamma(1+w^2)]m}{b w^2} \right. \\ 
    & \left. \quad \;\, + \, \frac{[w^4+\gamma^2(1-4w^2-w^4)]m^2}{2w^4b^2}-\frac{a^2}{2b^2}+\frac{2\gamma ma}{w b^2}+\mathcal{O}(\epsilon^3) \right \}.
\end{split}
\end{equation}
The calculation of the deflection angle of massive particles in the spacetime of rotating naked singularities requires integration of the first derivative of the azimuthal shift function $d\phi/dr$, as follows from Eq. (\ref{OrbitalEq}). Since, we are expecting small deviation in the orbital motion of the particle, we will consider only terms with the most significant contribution, for simplicity. Introducing new small enough parameters $\varepsilon_{m}=m/r_{0}$ and $\varepsilon_{a}=a/r_{0}$, we will perform power series expansion of $d\phi/dr$, in the spacetime under consideration (3), up to and including second order terms of $\varepsilon_{m}$ and $\varepsilon_{a}$. After the convenient change of variables $r=r_{0}/u$, we obtain
\begin{equation}\label{TaylorIFuncRNS}
\begin{split}
    \frac{d\phi}{du} \simeq \frac{1}{\sqrt{1-u^2}} & +\left[\frac{\gamma+w^2\left(u^2+u+\gamma-1\right)}{ w^2(1+u)\sqrt{1-u^2}}\right]\varepsilon_{m} +\frac{1}{2}\left[\frac{\mathcal{P}_{4}(u\,|\,\gamma,w)}{w^4(1+u)^2\sqrt{1-u^2}}\right]\varepsilon_{m}^2 \\
    & \pm\left[\frac{2 \gamma }{w(1+u) \sqrt{1-u^2}}\right]\varepsilon_{m}\varepsilon_{a} + \frac{1}{2}\left[\frac{1-2u^2}{\sqrt{1-u^2}}\right]\varepsilon_{a}^2 +\mathcal{O}(\varepsilon^3),
\end{split}
\end{equation}
where we are introducing the quartic polynomial in powers of $u$
\begin{equation}\label{F}
\begin{split}
    \mathcal{P}_{4}(u\,|\,\gamma,w) = & 3w^4u^4+6w^4u^3+(w^4+2w^4\gamma^2+6w^2\gamma^2)u^2 \nonumber \\
                & - [6w^4-6w^2(w^2+1)\gamma-(4w^2-4)\gamma^2]u + (w^4+4w^2-1)\gamma^2-w^4.
\end{split}
\end{equation}
Completing the integration of (\ref{TaylorIFuncRNS}) from the minimal distance of closest approach $r_{0}$ to observer and source positioned at infinite distance with respect to the center of the compact object, we obtain the following approximate expression for the deflection angle of a massive particle, according to Eq. (\ref{DA})
\begin{align}\label{DARNSro}
   \hat{\alpha}(r_{0}) & \simeq \frac{2m\gamma}{r_{0}w^{2}}(1+w^2)\pm\frac{4ma\gamma}{r_{0}^{2}w} \\
   & + \frac{m^2}{4 r_{0}^2 w^4}\Big{(}\pi w^2 \left[4\left(3+w^2\right)\gamma^2-w^2\right]+8 \gamma  \left(1+w^2\right) \left[w^2(1-\gamma)-\gamma\right]\Big{)}+\mathcal{O}(\varepsilon^3). \nonumber
\end{align}
The most significant contribution to the deflection angle of massive particles is given by massive 
term proportional to $\mathcal{O}(\varepsilon_{m})$, while the order proportional to angular momentum 
$\mathcal{O}(\varepsilon_{a})$ is vanishing. The next second order terms proportional to the mass 
$\mathcal{O}(\varepsilon_{m}^2)$ and to the mass and angular momentum $\mathcal{O}
(\varepsilon_{m}\varepsilon_{a})$ give also a contribution, as opposed to the second order term 
proportional to the square of the angular momentum $\mathcal{O}(\varepsilon_{a}^2)$, which is missing.
Moreover, taking into account the power series (\ref{ClosestDistanceRNS}), we can represent the 
deflection angle of the relativistic particles in terms of the impact parameter $b$ up to and including 
the second powers of the small parameters $\epsilon_{m}$ and $\epsilon_{a}$. After performing the 
calculations, we get the following invariant representation of the deflection for a given relativistic 
velocity $w$ of the massive particle
\begin{align}\label{DARNSb}
   \hat{\alpha}(b) & \simeq \frac{2m\gamma}{b w^{2}}(1+w^2) + \frac{m^2\gamma^2}{b^2}\left(1+\frac{3}{w^2}-\frac{1}{4\gamma^2}\right)\pi\pm\frac{4ma\gamma}{b^{2}w}+\mathcal{O}(\epsilon^3).
\end{align}
By comparison of the calculated approximate expression for the deflection angle of massive particles 
(\ref{DARNSb}) via geodesic approach with the result (\ref{DARNSGBT}) found by GBT it appears that up 
to the second formal order in the expansion parameters the both angles coincide.
Letting $w=1$, we recover the light deflection angle
\begin{equation}\label{DARNSLight}
    \hat{\alpha}(b) \simeq \frac{4m\gamma}{b}+\frac{4m^2\gamma^2}{b^2}\left(1-\frac{1}{16\gamma^2}\right)\pi\pm\frac{4ma\gamma}{b^2}+\mathcal{O}(\epsilon^3),
\end{equation}
presented in the static case in \cite{Gyulchev1, Virbhadra1, Virbhadra:2007kw}. Similarly, 
the plus (minus) sign represents the retrograde (prograde) light ray, respectively.

\subsection{Deflection angle by Kerr-like wormholes}

In the case of the Kerr-like wormhole (\ref{metric}) the impact parameter of the massive particles $\xi=wb$ with relativistic velocity $w$ and the minimal distance of closest approach $r_{0}$ obey the following relation
\begin{equation}\label{IParameterKLW}
    \xi(r_{0})=\frac{-2Ma+\sqrt{r_{0}(r_{0}^2+a^2-2Mr_{0})\big{[}r_{0}-(r_{0}-2M)(1-w^2)\big{]}}}{r_{0}-2M}
\end{equation}

Since, we are considering propagation of massive particles in the asymptotically flat spacetime regions with geometrical curvature, causing small deflection angles of trajectories, then $r_{0}$ is expected to be of the same order as the impact parameter $b$. Therefore, under assumptions $b\gg M$ and $b\gg a$, we are taking advantage of the two small enough expansion parameters \begin{equation}
    \epsilon_{M}=\frac{M}{b}, \quad\quad \epsilon_{a}=\frac{a}{b},
\end{equation}
where $M$ and $a$ are the mass and spin of the Kerr-like wormhole, respectively. Hence, we could find an approximate solution of the radial Eq. (\ref{RadialEqMotion}), as for itself we suggest the ansatz
\begin{equation}\label{r0approximation2}
    r_{0}\simeq b \left\{1+\sum\limits_{i,j=1}^{2} c_{M,a}\,\epsilon_{M}^{i}\epsilon_{a}^{j}+\mathcal{O}(\epsilon^3)\right\},
\end{equation}
Here the coefficients $c_{M,a}$ are real numbers.

Solving the equation $\dot{r}(r_{0})=0$, for the closest approach distance we obtain the following power series expansion
\begin{equation}\label{ClosestDistanceKLW}
\begin{split}
    r_{0}\simeq b \left\{1 -\frac{M}{b w^2}+\frac{(1-4w^2)M^2}{2w^4b^2}-\frac{a^2}{2b^2}+\frac{2Ma}{b^2 w}  +\mathcal{O}(\epsilon^3) \right\}.
\end{split}
\end{equation}

Now, we can proceed ahead toward finding of the perturbative expansion of the deflection angle of a massive particle moving in the spacetime of the Kerr-Like wormhole. Since, we are interested in the weak deflection limit, we will perform power series expansion of the derivative $d\phi/dr$, adopting the new small parameters $\varepsilon_{M}=M/r_{0}$ and $\varepsilon_{a}=a/r_{0}$ up to and including second order terms. In order to perform the integration afterwards more easily, we introduce a new variable $u=r_{0}/r$ in terms of which we obtain 
\begin{equation}\label{TaylorIFuncKLW}
\begin{split}
    \frac{d\phi}{du} \simeq \frac{1}{\sqrt{1-u^2}} & +\left[\frac{1+w^2(1+\lambda^2)(1+u)u}{w^2(1+u)\sqrt{1-u^2}}\right]\varepsilon_{m} +\frac{1}{2}\left[\frac{\mathcal{P}_{4}(u\,|\,\lambda,w)}{w^4(1+u)^2\sqrt{1-u^2}}\right]\varepsilon_{m}^2 \\
    & \pm\left[\frac{2}{w(1+u)\sqrt{1-u^2}}\right]\varepsilon_{m}\varepsilon_{a} + \frac{1}{2}\left[\frac{1-2u^2}{\sqrt{1-u^2}}\right]\varepsilon_{a}^2 +\mathcal{O}(\varepsilon^3),
\end{split}
\end{equation}
where it is introduced the fourth degree polynomial in variable $u$
\begin{equation}\label{FKLW}
\begin{split}
    \mathcal{P}_{4}(u\,|\,\lambda,w) = \; & 3w^4(1+\lambda^2)^2u^4+6w^4(1+\lambda^2)^2u^3 \nonumber \\
                & +w^2[3w^2(1+\lambda^2)^2+2(3+\lambda^2)]u^2 + 2[w^2(5+\lambda^2)-2]u + 4w^2-1.
\end{split}
\end{equation}
Therefore, the bending angle of the light ray, defined by Eq. (\ref{DA}), up to and including the 
second orders of the expansion parameters $\varepsilon_{M}$ and $\varepsilon_{a}$ is given by
\begin{align}\label{DAKLWro}
   \nonumber \hat{\alpha}(r_{0}) & \simeq \frac{2M}{r_{0}}\left(1+\lambda^2+\frac{1}{w^2}\right)\pm\frac{4Ma}{r_{0}^{2}}\frac{1}{w} \\
   & + \frac{M^2}{4 r_{0}^2 w^4}\Big{(}3\pi w^4(1+\lambda^2)^2+4w^2[3\pi-2+(\pi-2)\lambda^2]-8\Big{)}+\mathcal{O}(\varepsilon^3).
\end{align}
Taking advantage of the power series (\ref{ClosestDistanceKLW}) we can represent the deflection angle 
of the relativistic particles in terms of the impact parameter $b$, up to and including second order 
terms of the small parameters $\epsilon_{M}$ and $\epsilon_{a}$. Therefore, the deviation of 
trajectories of the massive particles by the Kerr-like wormhole leads to the following components of 
the deflection angle 
\begin{align}\label{DAKLWParticles}
    \hat{\alpha}(b)\simeq \frac{2M}{b}\left(1+\lambda^2+\frac{1}{w^2}\right) +\frac{M^2}{b^2}\left[\frac{3(1+\lambda^2)^2}{4}+\frac{3+\lambda^2}{w^2}\right]\pi \pm\frac{4Ma}{b^2}\frac{1}{w} +\mathcal{O}\left(\epsilon^3\right).
\end{align}
As in the previous case of the rotating naked singularities, here as well the most significant 
contribution to the deflection angle is given by the massive term proportional to $\mathcal{O}
(\epsilon_{M})$. The second order terms proportional to the mass $\mathcal{O}(\epsilon_{M}^2)$ and to 
the mass and angular momentum of the wormhole $\mathcal{O}(\epsilon_{M}\epsilon_{a})$ give a 
contribution to the deflection angle of the relativistic massive particles. The first and second order 
terms based on the wormhole's angular momentum, $\mathcal{O}(\epsilon_{a})$ and 
$\mathcal{O}(\epsilon_{a}^2)$ respectively, do not contribute in the perturbative analysis.
In the special case $w=c=1$, from expression (\ref{DAKLWParticles}) the deflection angle of light is recovered \cite{Ovgun:2018fnk}
\begin{align}\label{DAKLWLight}
    \hat{\alpha}(b)\simeq \frac{2M}{b}\left(2+\lambda^2\right) +\frac{M^2}{b^2}\left[\frac{3(1+\lambda^2)^2}{4}+\lambda^2+3\right]\pi \pm\frac{4Ma}{b^2} +\mathcal{O}\left(\epsilon^3\right).
\end{align}
Comparing between the obtained results for the deflection angles (\ref{DAKLWGBT}) and 
(\ref{DAKLWParticles}) via the GBT and geodesic approach respectively, shows coincidence of the 
perturbative expressions up to the leading terms proportional to $\mathcal{O}(\epsilon_{M}^2)$ and 
$\mathcal{O}(\epsilon_{M}\epsilon_{a})$.

\begin{figure}[h!]
\center\includegraphics[width=0.75\textwidth]{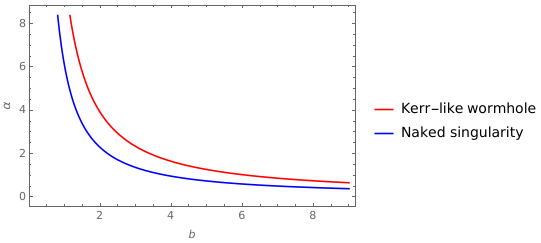}
\caption{The deflection angle as a function of $b$. We chose $\lambda=0.1$, $w=0.8$,$\gamma=0.6$, $m=M=a=1$.}
\label{w5}
\end{figure}

\section{Conclusions}
\label{concl}
We studied the gravitational deflection of massive/massless particles in a rotating 
naked singularities and Kerr-Like wormholes backgrounds.  In the first part of this paper, we used a recent geometric approach to compute
the gravitational deflection angle using the refractive index of the optical media and by utilizing the GBT applied to the optical 
metric. In order to incorporate the effect of relativistic speed on the deflection angle, we have assumed the propagating massive particles as a de Broglie wave packets yielding a 
modification of the refractive index $N$, and a constant quantity of a given optical media, $\lambda N=const$. In particular, 
we have shown that the refractive index and the Gaussian optical curvature are affected by the mass $M$, 
angular momentum parameter $a$, naked singularity parameter $\gamma$, the deformation parameter 
$\lambda$,  and the relativistic velocity of the particle $w$. Although the geodesic curvature $\kappa(C_R)$ is 
slightly modified due the velocity of the particle, i.e. $\kappa(C_R) \to w R^{-1}$,  we have shown that the expression for the deflection angle reads
\begin{equation}\notag
\hat{\alpha}=-\int\limits_{0}^{\pi}\int\limits_{r_{\gamma}}^{\infty}\,\mathcal{K}\,dS.
\end{equation}

In other words we have found exactly the same expression as in the case of deflection of light, which suggests that the above result can be viewed as a form-invariant quantity for
asymptotically flat spacetimes. Again, the total deflection angle is found by integrating over a domain outside the light ray 
described by $r_{\gamma}$, hence the global effects are important. Note that in our setup we have 
simplified the problem by considering a linearized spacetime metrics and we have also assumed that the source and the observer are located at infinity. Applying then the GBT 
to the isotropic metrics we have found two interesting results for the deflection angles, namely
\begin{equation}\notag
\hat{\alpha}_{Naked\, sing.}\simeq \frac{2 m \gamma}{b w^2}(1+w^2) \pm \frac{4 m a \gamma}{b^2 w}
\end{equation}
in the case of rotating naked singularity, and
\begin{equation}\notag
\hat{\alpha}_{Kerr-like \,wormhole}\simeq \frac{2M}{b}\left(1+\lambda^2+\frac{1}{w^2}\right)\pm \frac{4Ma}{b^2}\frac{1}{w}
\end{equation}
in the case of Kerr-like wormholes. From the last two equations we see that an apparent singularity appears when $w\to 0$, therefore an additional constraint should be imposed, namely $0 < w \leq 1$. Now we specialize further the case of $\lambda=0$ and $\gamma=1$, in that case both results coincide with the Kerr deflection angle for massive particles reported recently in  \cite{Jusufi:2018kry}, but also in \cite{g1,g2}. Note that our approach works perfectly well for relativistic particles, in this sense the apparent singularity is not problematic.  Furthermore, we did a careful analyses by studying the problem of deflection of particles using the Hamilton-Jacobi equation and we find agreement of results in leading order terms. Our results show that, while the deflection angle decreases with $\gamma$ in the case of JNW spacetime, contrary to this, the deflection angle increases with the deformation parameter $\lambda^2$ in the case of Kerr-like wormholes as can be seen from figure \eqref{w5}. Our results are not important only from conceptual point of view, namely the difference between deflection angels may shed light from observational point of view in order to distinguish the two spacetimes. Finally,  we plan in the near future to extend these results by including finite distance corrections on the deflection angle, in other words by assuming that the distance between the source and the observer is finite.

\textbf{Acknowledgments}: We would like to thank the anonymous reviewer for enlightening comments related
to this work.

\end{document}